\documentclass{czjphyse}

\usepackage{cite}
\usepackage{epsf}
\begin{document}
\title{RADIATIVE CORRECTIONS TO THE STRUCTURE FUNCTIONS AND SUM RULES IN
POLARIZED DIS } 
\authori{I. Akushevich, A. Ilyichev, N. Shumeiko }
\addressi{National Centre for  Particle and  High Energy Physics,
Bogdanovich str. 153, 220040 Minsk, Belarus}
\authorii{}     
\addressii{}    
\authoriii{}    
\addressiii{}   
\headtitle{ Radiative corrections to the structure functions and
sum rules in
polarized DIS \ldots} 
\headauthor{ I.Akushevich et al.}
\specialhead{ I.Akushevich et al:Radiative corrections to the structure
functions and sum rules in polarized DIS\ldots}
\evidence{A}
\daterec{XXX}    
\cislo{0}  \year{1999}
\setcounter{page}{1}
\pagesfromto{000--000}
\maketitle

\begin{abstract}

The one-loop NLO radiative corrections (RC) to the observables in
polarized DIS using assumption that a quark is an essential massive
particle are considered. If compared with classical QCD formulae the
obtained results are identical for the unpolarized and different for
polarized sum rules, that can be explained as the influence of the finite
quark mass effects on NLO QCD corrections. The explicit expression for
one-loop NLO QCD contribution to the structure function $g_2$ is
presented.

\end{abstract}

\section{Introduction}

\begin{figure}
\centering
\vspace{0.5cm}
\hspace{-7cm}
\unitlength 1mm
\begin{picture}(100,80)
\put(10,0){
\epsfxsize=8cm
\epsfysize=8cm
\epsfbox{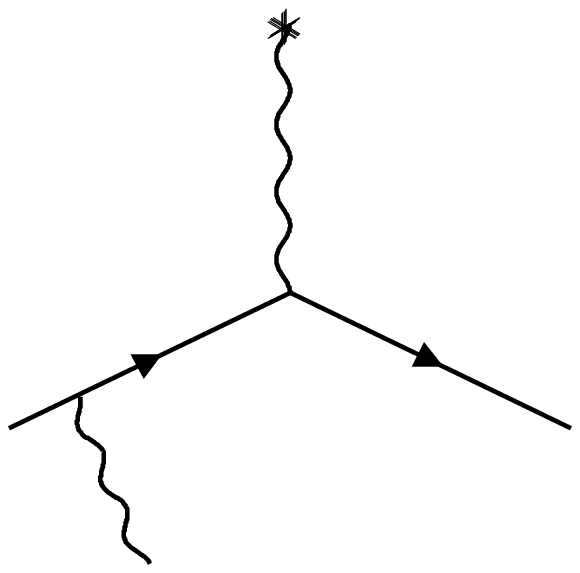}
}
\put(50,0){
\epsfxsize=8cm
\epsfysize=8cm
\epsfbox{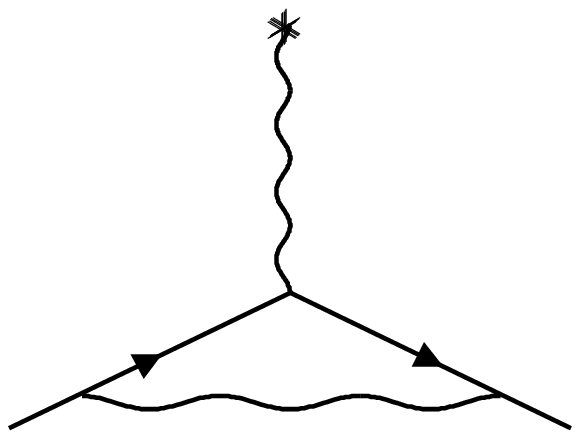}
}
\put(90,0){
\epsfxsize=8cm
\epsfysize=8cm
\epsfbox{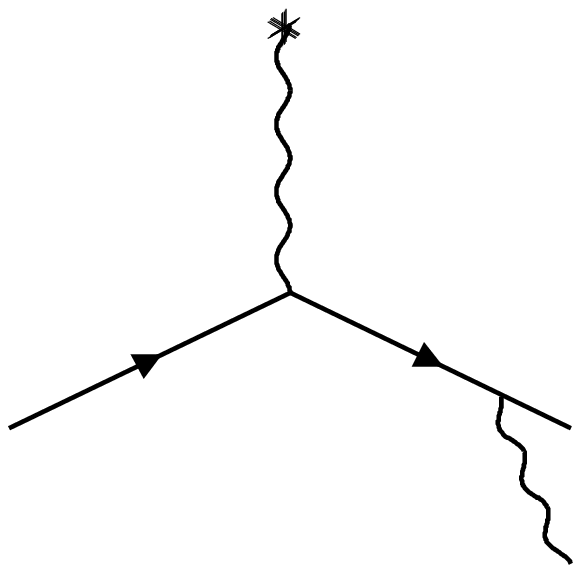}
}
\end{picture}
\vspace*{-4.5cm}
\caption{Full set of Feynman graphs for one-loop QCD or QED correction
to the hadronic current}
\label{F1}
\end{figure}

Notice that QCD and QED RC having different origins possess on the
one-loop level some common features. If to restrict our consideration to
so-called QCD Compton process then both these corrections will be
described by the identical set of Feynman graphs (see fig.\ref{F1}).
Nevertheless it has to be emphasized that different methods of
calculations are used within the framework of these theories.

One of the standard approach to the calculation of electromagnetic RC to
the hadronic current is based on the cancellation of the infrared
divergences by Bardin-Shumeiko method \cite{BS,ABK} developed to the case
of
polarized particles in refs.\cite{KSh,ASh}. Later this approach will be
called as the "massive" scheme that supposes the presence of a quark as an
essential massive particle.

At the same time the "classical" calculations of RC in the perturbative
QCD are performed in the chiral limit {\it i.e.} when a quark mass is
equal to zero \cite{AGL,AEL,GRV2,QCD}. Later that method of
calculation will be called as the "massless" scheme. However there is no
fundamental reason to consider QCD RC within the "massless" scheme
especially when we deal with such fine effects as polarized ones.


As an example let us consider RC to the lepton current in polarized DIS.
The polarized part of the cross section of the process is proportional to
the polarization vector of the scattering lepton that has the form
\cite{ASh}:
\begin{equation} 
\xi _L=\frac {k} {m}-m\frac {p}{pk},
\label{xii} 
\end{equation}
where $k$ ($p$) is a momentum of an initial  lepton (target). As
a rule the
second part of the polarization vector can be dropped. But, as it was
shown in \cite{AISh,BARD,LRC}, when the radiation of a real photon is
considered even in NLO approximation the second term of (\ref{xii}) gives
non-zero contribution to the polarized observables. It has to be noted
that the same situation appears when QED RC to the hadronic current
\cite{AISh} are calculated and has to be expected in QCD.


Unfortunately the "massless" scheme can not take into account the second
term in the polarization vector of the target by definition. So the main
aim of this report is application of the "massive" scheme usually used in
QED to the calculation of QCD RC within a naive quark-parton model in
order to estimate the value of the finite quark mass effect in polarized
DIS.


\section{Method of Calculation and Main Results}

The lowest-order one-loop 
RC to the hadronic current consist of two parts
whose contribution are calculated in a different way.
$$
W^{1-loop}_{\mu \nu}=
W^R_{\mu \nu}+
W^V_{\mu \nu}.
$$

The first one appears from a gluon emission and requires the integration
over its phase space:
\begin{eqnarray}
W^R_{\mu \nu} = 
\frac {\alpha_s}{12\pi ^2}
\sum _q e_q^2  \int
\frac {d^3k}{k_0}\frac 1 {(p_{2q}-p_{1q})^2}
\Biggl [   
Sp \; \Gamma _{\mu \alpha}
(\hat {p}_{1q}+m_q)
\bar{\Gamma } _{\alpha \nu}(\hat {p}_{2q}+m_q)f _q
\nonumber \\ \qquad\qquad
+Sp \; \Gamma _{\mu \alpha}
(\hat {p}_{1q}+m_q)
\gamma_5 \hat\eta
\bar{\Gamma } _{\alpha \nu}(\hat {p}_{2q}+m_q)\Delta f _q
\Biggl ],
\label {wra}
\end {eqnarray}
where $p_{1q}$ ($p_{2q}$) in an initial (final) 4-momentum of the quark
and
$$ 
\Gamma _{\mu \alpha}=
2\Omega _q^{\alpha }
\gamma_{\mu}-   
\frac {\gamma_{\mu}\hat{k}\gamma_{\alpha}}{2kp_{1q}}-
\frac{\gamma_{\alpha}\hat{k}\gamma_{\mu}}{2kp_{2q}}
,\qquad
\bar{\Gamma} _{\alpha \nu}=
2\Omega _q^{\alpha }
\gamma_{\nu}
-\frac{\gamma_{\alpha}\hat{k}\gamma_{\nu}}{2kp_{1q}}
-\frac {\gamma_{\nu}\hat{k}\gamma_{\alpha}}{2kp_{2q}},
$$
$$
\Omega _q=
\frac {p_{1q}}{2kp_{1q}}- \frac {p_{2q}}{2kp_{2q}}.
$$

The second part  appearing from a
gluon exchange graph reads 
\begin{equation}
W_{\mu \nu}^V=\frac 43 \frac {\alpha _s}{\pi}
\sum _q \biggl [-2\left ({\cal P}^{IR}+\log \frac {m_q}{\mu} \right 
)(l_q-1)
 -\frac 12 l_q^2
+\frac 32 l_q-2+\frac{\pi ^2}6\biggl ]
W_{\mu \nu}^{0q}.
\end{equation}
where $W^{0q}_{\mu \nu}$ is a contribution of $q$-quark to  the hadronic
tensor on the Born level. 
The pole term which corresponds to the infrared divergence is contained in
${\cal P}^{IR}$. The arbitrary parameter $\mu $ has a dimension of a
mass.

 Both of these contributions include the infrared divergences, which have
to be careful considered in order to be canceled. Like QED we use the
identity:
$$
W^R_{\mu \nu }=W^R_{\mu \nu }-W^{IR}_{\mu \nu }+W^{IR}_{\mu \nu }=
W^{F}_{\mu \nu }+W^{IR}_{\mu \nu }.
$$
Here $W^{F}_{\mu \nu }$ is finite for $k \rightarrow 0$, and
$W^{IR}_{\mu
\nu }$ is
the infrared divergent part of (\ref{wra}). Using the dimensional
regularization scheme the latter can be given in the form
\begin{eqnarray}
W^{IR}_{\mu \nu }&=&\frac 4 3 \frac {\alpha }{\pi }\sum _q
\biggl[2\biggl ({\cal P}^{IR}+\log \frac {m_q}{\mu} \biggr)(l_q-1)
+l_ql_v+\frac 12 l_q^2 \qquad
\nonumber \\ &&\qquad\qquad\qquad
-\frac 12 l_v^2
-\frac 34 l_q-\frac 74 l_v+\frac 34 -\frac{\pi ^2}3\biggr]
W^{0q}_{\mu
\nu },
\nonumber
\end{eqnarray}
where
$$
l_q=\log \frac{Q^2}{m_q^2}, \qquad
l_v=\log \frac {1-x}x.
$$

The sum of $W_{\mu \nu}^{IR}$ and
 $W_{\mu \nu}^V$ 
\begin{equation}
W_{\mu \nu}^{IR}+
W_{\mu \nu}^V
=\frac  23 \frac {\alpha _s}{\pi}\sum _q
\bigl[2l_ql_v-l_v^2+\frac 32 l_q
-\frac 72 l_v-\frac 52
-\frac {\pi^2}{3}
\bigr]W_{\mu \nu}^{0q}
=\frac 23 \frac {\alpha _s}{\pi}\sum _q\delta _qW_{\mu \nu}^{0q}
\label{delq}
\end{equation}
is infrared free.

In order to extract some information about QCD contribution to the
polarized structure function $g_2$ the integration in $W^F_{\mu \nu}$
over the gluon phase space should be performed without any assumptions
about the polarization vector $\eta $. So the technique of tensor
integration have to be applied in this case. Since the result of the
analytical integration has the same tensor structure as the usual hadronic
tensor in polarized DIS, the coefficients in front of the corresponding
tensor structures (like $g_{\mu\nu}$, $p_\mu p_\nu$ ... ) can be
interpreted as one-loop QCD contributions to the corresponding structure
functions.

Thus the QCD-improved structure functions read:
\begin{eqnarray}
F_1(x,Q^2)&=&
\frac 1{2x}[F_2(x,Q^2)-F_L(x,Q^2)],
\nonumber \\
F_2(x,Q^2)&=&
x \sum _q e_q^2f_q(x,Q^2),
\nonumber \\
F_L(x, Q^2)&=&
\frac {4\alpha _s}{3\pi}x\sum _q e_q^2\int
\limits_x^1dzf_q(x/z),
\nonumber \\
g_1(x,Q^2)&=&
\frac 12 \sum _q e_q^2\Delta f_q(x,Q^2),
\nonumber \\
g_2(x,Q^2)&=&
-\frac {\alpha _s}{6\pi}\sum _q e_q^2
\biggl\{
(l_q+\log (1-x)
-1)\Delta f _q(x)
+\int \limits ^1 _xdz
\Bigl [
(4l_q
\nonumber \\
&&-4\log z(1-z)
+12
+\frac 1{(1-z)})\Delta f_q(x/z)
-\frac{\Delta f_q(x)}{(1-z)}
\Bigl ]
\biggl\},
\nonumber
\end {eqnarray}
where the $Q^2$-dependent unpolarized and polarized parton distributions
are defined as
\begin{eqnarray}
f_q(x, Q^2)=
(1+\frac {2\alpha _s}{3\pi} \delta _q)
f_q(x)
+\frac {2\alpha _s}{3\pi} \int\limits_x^1 \frac{dz}{z}
\Bigl[
(\frac {1+z^2}{1-z}
( l_q
-\log z(1-z))
\qquad
\nonumber \\
-\frac 72 \frac 1{1-z}
+3z
+4)f_q(\frac{x}{z})
-\frac 2{1-z}\left(l_q+\log \frac{z}{1-z}
-\frac 74\right)f_q(x)
\Bigl],
\nonumber \\
\Delta f_q(x,Q^2)=
(1+\frac {2\alpha _s}{3\pi} \delta _q)
\Delta f_q(x)
+ \frac {2\alpha _s}{3\pi} \int\limits_x^1 \frac{dz}{z}
\Bigl[(\frac{1+z^2}{1-z}
(l_q
-\log z(1-z))
\qquad
\nonumber \\
-\frac 72 \frac 1{1-z}
+4z
+1)\Delta f _q(\frac{x}{z})
-\frac 2{1-z}\left(l_q+\log \frac{z}{1-z}
-\frac 74\right)\Delta f _q(x)
\Bigl],
\nonumber
\end{eqnarray}
and $\delta _q $ can be found in (\ref{delq}).

Now it is interesting to compare the one-loop QCD contribution to the
structure functions presented above with the classical results obtained
within massless scheme.

It is clear that the explicit expressions found in both of the discussed
schemes have the same structure.  However the results obtained by us in
spite of the standard ones do not require any renormalization (see
\cite{QCD} for details). The another interesting issue of our approach is
the explicit finite expression for $g_2$, which cannot be obtained in the
"massless" scheme since the first moment of $g_2$ includes the divergence.


After integration of the QCD-improved structure functions over the
scaling
variable $x$ the explicit QCD contributions to the sum rules can be
 presented by the following expressions:  
\begin{eqnarray}
\int \limits _0^1 dx F_1(x,Q^2)&=&
(1+C_{f1}\frac {\alpha_s}{\pi})
\int \limits _0^1 dx F_1^0(x),
\nonumber \\
\int \limits _0^1 \frac {dx}{x} F_2(x,Q^2)&=&
(1+C_{f2}\frac {\alpha_s}{\pi})
\int \limits _0^1 \frac {dx}{x} F_2^0(x),
\nonumber \\
\int \limits _0^1 dx g_1(x,Q^2)&=&
(1+C_{g1}\frac {\alpha_s}{\pi})
\int \limits _0^1 dx g_1^0(x),
\nonumber \\
\int \limits _0^1 dx g_2(x,Q^2) &=& \frac {\alpha _s}{\pi}\sum _q e^2_q
C_{g2}
\int \limits _0^1 dx \Delta f_q(x),
\nonumber
\end{eqnarray}
where the structure functions with index "0" 
are defined in the naive parton model, and the coefficient
$C_{f1,f2,g1,g2}$ both to the well-known
classical ($m_q=0$) and our results ($m_q\ne 0$) in the table \ref{tab1} 
are presented.
\begin{table}[t]
\begin{center}
\begin{tabular}{c|c|c|c|c}
\hline
&
$C_{f1}$&
$C_{f2}$&
$C_{g1}$&
$C_{g2}$
\\
\hline &&&
\\
[-0.2cm]
$m_q=0$&
$-2/3$&
$0$&
$-1$&
$\infty $
\\
$m_q\ne 0$&
$-2/3$&
$0$&
$-5/3$&
$-l_q/2-4/3$
\\
[0.3cm]
\hline
\end{tabular}
\caption{Corrections to sum rules in polarized DIS}
\label{tab1}
\end{center}
\end{table}

Notice that in the both of the discussed schemes the QCD corrections to
the unpolarized sum rules have the identical values. However the
difference between the QCD-corrections to the polarized sum rules are
existed. The origin of the disagreement for $C_{g1}$ can be visualized in
the case of longitudinal polarized DIS: the additional contribution comes
from that part of the polarization vector which is proportional to the
quark mass. The similar situation for QED corrections to the leptonic
current has been already discussed above. The different value of the
$C_{g2}$ can be explained in a simple way: $l_q \rightarrow \infty $ when
$m_q \rightarrow 0$.

As it was mentioned above our calculation was performed within a naive
parton model when the quark mass is defined as $m_q=xM$.  Another
possibility is to consider it as a constant. However an additional
contribution appearing during the integration of the QCD-improved
structure functions over $x$ is completely cancels due to DGLAP equations
and the final results for these two cases are identical.

The additional contributions to the spin dependent sum rules influence
on the experimentally measured first moment of $g_1$

$$
\Gamma ^{meas}_1=(1+C_{g1}\frac {\alpha_s}{\pi})\Gamma ^0_1,
$$
where $\Gamma ^{meas}_1$ and $\Gamma ^0_1 $ are  measured and
extracted
quantities respectively.
It could be seen from the follow relation:
$$
\Gamma ^0_{1\;m_q\ne 0}/
\Gamma ^0_{1\;m_q=0}
=
(1-\frac {\alpha_s}{\pi})/
(1-\frac {5\alpha_s}{3\pi})\sim 1.07,
$$
that the influence of the finite quark mass on QCD RC to the first moment
of $g_1$
is approximately $7\% $ for $\alpha_s=0.27$.

Note that within our approach we can estimate the first moment of
$g_2$. As can be seen from the table \ref{tab1}
the sign of the first moment of $g_2$
calculated within our scheme
$$
\int\limits^1 _0 dx g_2< 0,
$$
is in  agreement  with the result of SLAC experiment \cite{SLAC}.

\end{document}